\documentclass[preprint,unsortedaddress,superscriptaddress,footinbib,amsmath,amssymb,aps,pra,showkeys,showpacs]{revtex4-2}
\usepackage{graphicx}
\usepackage{xcolor}
\usepackage{amsmath}
\usepackage{amssymb}
\usepackage{float}
\usepackage{braket}
\begin{document}

\title{Schr\"odinger cats coupled with cavities losses: the effect of finite and structured reservoirs.}

\author{J. Lira}
\email{jeffyfisica@gmail.com}
\affiliation{Colegiado de Engenharia, Unex, 45020-510, Vit\'oria da Conquista-BA, Brazil}
\author{L. Sanz}
\email{lsanz@ufu.br}
\affiliation{Instituto de F\'{i}sica, Universidade Federal de Uberl\^andia,38400-902, Minas Gerais, Brazil}

\begin{abstract}
We discuss the generation of a Schr\"odinger cat in a nanocavity created by the coupling of an electromagnetic mode with an exciton in a quantum dot considering the dispersive limit of the Jaynes-Cummings model. More than the generation itself, we focus on the effects of the environment over the bosonic state in the nanocavity, which has losses simulated by coupling with two different kind of reservoirs. In the first case, the interaction between the system with a finite reservoir shows that fragments of different sizes of the reservoir deliver the same amount of information about the physical system in the dynamics of the birth and death of the Schr\"odinger cat. The second case considers a structured reservoir, whose spectral density varies significantly with frequency. This situation becomes relevant in solid-state devices where quantum channels are embedded, as memory effects generally cannot be neglected. Under these circumstances, it is observed that the dynamics can differ substantially from the Markovian, presenting oscillations related to the average number of photons. These oscillations influence the information flow between the system and the environment, evidenced here by the measurement of non-Markovianity.
\end{abstract}
\maketitle
\section{Introduction}
\label{sec:intro}

\hspace{10pt}Since the beginning of the quantum theory, the Schr\"odinger cat state~\cite{Schrodinger1935,Gerry97,RevModPhys.85.1083,RevModPhys.85.1103} has been pointed out as one of the crossing points between both, the classical and the quantum worlds. Successful experimental implementations have included several studies on cavity quantum electrodynamics involving Rydberg atoms passing through superconducting cavities~\cite{Deleglise08}, cold ions trapped within cavities~\cite{Monroe1131}, and even a transmon qubit coupled to cavities in a circuit quantum electrodynamics system~\cite{Haroche2020,Sun2014}. In the initial experimental setup, for example, the state of the cavity must be probed through the subsequent passage of new atoms, while the cavity may simultaneously lose photons. The experiments also offer the possibility of reconstructing the Wigner function~\cite{Dowling94,Hillery84,Bertet02,Deleglise08}. A particular instance of these cat states, specifically the one proposed by Yurke and Stoler~\cite{YurkeStolerCat}, has proven difficult to implement experimentally due to its precise requirement for a relative phase ($\pm i$) between the superposition states. The average number of atoms in these cavities is low (approximately 4), hence the term ``kittens" rather than ``cats" is used to describe the state of the cavity. 

Regarding the boundary between quantum and classical mechanics, it is interesting to explore how a quantum state, once generated, evolves under the influence of the decoherence process~\cite{Breuerbook,Shishkov_2019,zurek2003}. Recent studies have demonstrated that quantum information reaching the environment is not always irretrievably lost~\cite{rivas2014,breuer2016,de2017}; sometimes, it can be recovered, at least temporarily. For instance, a team from the Federal University of Rio de Janeiro - UFRJ conducted an experiment with photons, demonstrating that under certain conditions, some information is preserved and might even be retrievable~\cite{farias2012}. Consistent with this finding, an entropic measure has been established as an indicator of the correlation between non-Markovian behavior and the information exchange between the system and its environment~\cite{fanchini2014}. In this particular case, a two-level system undergoing a relaxation process was examined, utilizing an experimental optical setup that provides complete access to the environmental state. To enhance our comprehension of how a system interacts with its environment, we can ``engineer" the environment. This engineered approach allows us to characterize, control, and possibly reduce the adverse effects on the system~\cite{Harrington2022}.

Some researchers contend that merely stating decoherence drives quantum behavior, thus rendering it classical to an observer, is insufficient to explain the emergence of classical, objective reality~\cite{darwnismorevista}. There must be a consensus among multiple observers regarding the properties of quantum systems. To achieve agreement on a measured value (a hallmark of classicality) numerous replications are necessary. For instance, if ten observers measure the position of a speck of dust, they will find it in the same location because each observer has access to a distinct replica of the information~\cite{zurek2003}. In this perspective, {monitoring a small portion of the environment is sufficient to obtain most of the available information}, and observing more than a fraction yields diminishing returns. A prominent concept within this theoretical framework is that the defined properties of objects, which we associate with classical physics, emerge from a range of quantum possibilities through a process somewhat akin to natural selection in evolution. The properties that persist are, in a sense, the fittest. 

In this study, we examine the interaction between a two-level system illustrated by an exciton in a quantum dot and a bosonic mode within a cavity, similar to the nanocavities that interact with excitons as described in Ref.~\cite{Freitas17}. This interaction experiences energy dissipation and is characterized by a Hamiltonian that models the Jaynes-Cummings interaction under the dispersive approximation. The dissipating cavity mode is depicted as a harmonic oscillator linked linearly to $N$ oscillators, which represent the reservoir~\cite{de2014,oliveira2019}. In the field of condensed-matter physics, the production of quantum light using quantum dots within nanocavities has been extensively discussed~\cite{Muller15,Heinze15,Munoz14,Farrow08,Kiraz04}. However, the feasibility of producing a Schr\"odinger cat state with the specific system has not been studied, although the key element, the dispersive interaction between a two-level system and a boson state, is present in this experimental context. We will focus our attention on the reservoir, analyzing the system interaction with two types of reservoirs, one finite and the other structured, conducting an analysis of dynamics of the normalized average mutual information, entanglement measures and non-Markovianity in our system.

This paper is structured as follows: Section~\ref{sec:model} introduces the model of our physical system, along with some essential definitions and mathematical tools utilized in our calculations. Section~\ref{sec:results} is divided into two subsections; the first offers a general discussion on the quantum dynamics of the system, and the second details our principal findings concerning the effects of two types of reservoirs, finite and structured, and the behavior of quantum information in these non-trivial scenarios. Lastly, Section~\ref{sec:summary} provides a summary of our work.

\section{Physical System, model and mathematical tools.}
\label{sec:model}

\hspace{10pt}Our system of interest is composed of a two-level particle or quasi-particle interacting with bosons, as modeled by the Jaynes-Cummings model. We consider an exciton in a quantum dot, which acts as the two-level system, encoding a qubit, while the electromagnetic mode represents the single mode of a photonic nanocavity. To encode the qubit, we define $\lvert 0\rangle$ as the state without an exciton and $\lvert 1 \rangle$ for the direct exciton state. Conversely, to generate and sustain a coherent state within the nanocavity, one can employ a continuous laser resonant with the cavity mode as described in Ref.~\cite{Freitas17}. Initially, for qualitative analysis, we will omit this source of cavity pumping.

The system's Hamiltonian is expressed as follows: ($\hbar=1$)
\begin{equation}
\label{eq:hamiltonian}
\hat{H}=\omega_x\hat{\sigma}_+\hat{\sigma}_-+\omega_c\hat{a}^{\dagger}\hat{a}+g(\hat{\sigma}_+\hat{a}+\hat{\sigma}_-\hat{a}^{\dagger})\ ,
\end{equation} 
where $\hat{\sigma}_{\pm}$ are the pseudospin operators that describe the exciton, and $\hat{a}^{\dagger}$ ($\hat{a}$) is the creation (annihilation) operator of the cavity mode. 
The frequencies of the exciton and cavity mode are denoted by $\omega_x$ and $\omega_c$, respectively, and $g$ signifies the exciton-cavity coupling strength.

When the frequency of the exciton is significantly different from that of the cavity mode, the system is considered to be in a dispersive regime. The interaction between an exciton and a cavity mode under the dispersive approximation is described by an effective Hamiltonian. This effective Hamiltonian, as detailed in Ref.~\cite{PhysRevA.59.3918}, is presented as follows:
\begin{equation}\label{hamiltoniano efetivo}
\hat{H}_{\mathrm{eff}}=\omega_x\hat{\sigma}_+\hat{\sigma}_- +\omega_c\hat{a}^{\dagger}\hat{a}+\omega\hat{a}^{\dagger}\hat{a}\sigma_z\ .
\end{equation}

At this point, we are ready to make some considerations about the losses in the cavity. Instead of consider the usual master equation considering Lindblad operators, it is our interest to consider that the cavity mode suffers dissipation by a bath of $N$ coupled linear oscilators. In that way, we want to evaluate the effects of reservoirs over the dynamics of our now open quantum system. The Hamiltonian reads as:
\begin{eqnarray}\label{hamiltoniano t}
H_t =  \hat{H}_{\mathrm{eff}}+ \sum_{k}\omega_k\hat{b}_k^{\dagger}\hat{b}_k + \sum_{k}\gamma_k(\hat{a}^{\dagger}\hat{b}_k + \hat{a}\hat{b}_k^{\dagger}),
\end{eqnarray}  
where $\hat{b}_k$ is the $k-$th operator on the bosonic bath with frequency $\omega_k$ and $\gamma_k$ is the coupling constant between the mode in the nanocavity and the same mode. In principle, the oscilators in the bath are not interacting. Still, it is expected an effective interaction mediated by the nanocavity.

Once we are considering the whole system, the dynamics is governed by the Schr\"odinger equation:
\begin{eqnarray}\label{eq schrodinger}
H\lvert \psi(t)\rangle = i\frac{d}{dt}\lvert \psi(t)\rangle.
\end{eqnarray}

A mathematical tool that we will use in some of our calculations is the encoding of a superposition state of coherent states using a basis of two bosonic qubits. Consider a generic quantum state as given by:
\begin{equation}
    \lvert \psi \rangle=a\lvert \alpha ,\lambda \rangle+b\lvert \beta, \chi \rangle,
\end{equation}
where $\left\{\lvert\alpha\rangle,\lvert\beta\rangle\right\}$ belongs to a subsystem $A$ and $\left\{\lvert\lambda\rangle,\lvert\chi\rangle\right\}$ belongs to a subsystem $B$. The objective is to express this state as a generic superposition within the bosonic basis $\left\{\lvert 0\rangle,\lvert 1\rangle\right\}_A\otimes\left\{\lvert 0\rangle,\lvert 1\rangle\right\}_B$ written as
\begin{equation}
    \lvert \psi \rangle=p_+\lvert 11\rangle_{AB}+q_+\lvert 10\rangle_{AB}+q_-\lvert 01\rangle_{AB}+p_-\lvert 00\rangle_{AB},
\label{cod_bosonicQB}
\end{equation}
where one need to find the expressions for the four unknown parameters $p_{\pm}$ and $q_{\pm}$. The reader is invited to verify that this is accomplished by applying the subsequent transformations:
\begin{eqnarray}
\label{eq:transformation2qubits}
    \lvert\alpha\rangle&=&S_+\lvert 1\rangle_A - S_-\lvert 0\rangle_A,\nonumber\\
    \lvert\beta\rangle&=&e^{-i\theta}\left(S_+\lvert 1\rangle_A + S_-\lvert 0\rangle_A\right),\nonumber\\
    \lvert\lambda\rangle&=&R_+\lvert 1\rangle_B - R_-\lvert 0\rangle_B,\nonumber\\
    \lvert\chi\rangle&=&e^{-i\phi}\left(R_+\lvert 1\rangle_B + R_-\lvert 0\rangle_B\right),
\end{eqnarray} 
where we define
\begin{equation}
    S_{\pm}=\sqrt{\frac{1\pm\lvert\langle \alpha\vert\beta\rangle\rvert}{2}},\, R_{\pm}=\sqrt{\frac{1\pm\lvert\langle \lambda\vert\chi\rangle\rvert}{2}},\,e^{i\theta}=\frac{\langle\beta\vert\alpha\rangle}{\lvert\langle\alpha\vert\beta\rangle\rvert},\,e^{i\phi}=\frac{\langle\chi\vert\lambda\rangle}{\lvert\langle\lambda\vert\chi\rangle\rvert}.
\end{equation}
resulting in the coefficients in Eq.~(\ref{cod_bosonicQB}) given by:
\begin{equation}
    p_{\pm}=\left[a+be^{-i\left(\theta+\phi\right)}\right]S_{\pm}R_{\pm}, q_{\pm}=\left[-a+be^{-i\left(\theta+\phi\right)}\right]S_{\pm}R_{\mp}
\end{equation}

\section{Results}
\label{sec:results}
\subsection{Dynamics of the physical system}
\label{subsec:dynamics}
\hspace{10pt}Consider the physical system set in the initial state given by 
\begin{equation}
\label{estado incial t}
\lvert\psi(0)\rangle =(\sin\varphi\lvert 1\rangle +\cos\varphi\lvert 0\rangle )\otimes\lvert\alpha\rangle\otimes\prod_{k}\lvert 0_k\rangle \ ,
\end{equation}

At this point, we can assert that the hamiltonian (\ref{hamiltoniano t}) evolves this initial state into a superposition of coherent states in both the cavity and the reservoirs with undetermined amplitudes $\alpha(t)$, $\beta(t)$, $\lambda(t)$, and $\chi(t)$ as follows:

\begin{equation}
\label{estado evoluido t}
\lvert\psi(t)\rangle =\sin\varphi\lvert 1\rangle\otimes\lvert \alpha(t)\rangle\otimes\lvert \lambda(t)\rangle +\cos\varphi\lvert 0\rangle\otimes\lvert\beta(t)\rangle\otimes\lvert \chi(t)\rangle=\lvert\psi(t)\rangle_1+\lvert\psi(t)\rangle_2 ,
\end{equation}
where $\lvert \lambda(t)\rangle = \prod_{k}^{}\lvert \lambda_k(t)\rangle$ e $\lvert\chi(t)\rangle = \prod_{k}^{}\lvert \chi_k(t)\rangle$, with $\alpha(t)$ ($\beta(t)$) and $\lambda(t)$ ($\chi(t)$) being the amplitudes of the coherent states of the nanocavity and the reservoir, respectively.
The state $\lvert\psi(t)\rangle$ must be a solution of the Schr\"odinger equation (\ref{eq schrodinger}) which implies that we will obtain a set of coupled differential equations for the amplitudes $\alpha(t)$ and $\lambda(t)$, associated with the first term in Eq.(\ref{estado evoluido t}) as well as $\beta(t)$ and $\chi(t)$, corresponding to the second term. 

Let us work with the first part. We start with the equation:
\begin{eqnarray}\label{psi1}
\frac{d}{dt}\lvert \psi(t)\rangle_1 = \frac{d}{dt}(\sin\varphi\lvert 1\rangle\otimes\lvert\alpha(t)\rangle\otimes\lvert \lambda(t)\rangle). 
\end{eqnarray}
For a generic coherent state $\lvert\xi\rangle$, it is true that
\begin{equation}
\frac{d}{dt}\lvert\xi(t)\rangle =\frac{d}{dt}\Big(e^{-|\xi(t)|^2/2}\sum_{n=0}^{\infty}\frac{\xi(t)^n}{\sqrt{n!}}\lvert n\rangle \Big)
=-\frac{1}{2}\Big(\frac{d}{dt}\lvert\xi(t)\rvert^2\Big)\lvert\xi(t)\rangle + \dot{\xi}(t)\hat{a}^{\dagger}\lvert\xi(t)\rangle\,
\end{equation}
thus we can rewrite Eq.(\ref{psi1}) as
\begin{eqnarray}
\frac{d}{dt}\lvert   \psi(t)\rangle_1 = \bigg[-\frac{1}{2}\frac{d}{dt}\Big(|\alpha(t)|^2 + \sum_{k}^{}|\lambda_k(t)|^2\Big) + \dot{\alpha}(t)\hat{a}^{\dagger} + \sum_{k}^{}\dot{\lambda}_k(t)\hat{b}_k^{\dagger}\bigg]\lvert  \psi(t)\rangle_1\ , 
\end{eqnarray}
where the total excitation number is a constant given by $|\alpha(t)|^2 + \sum_{k}^{}|\lambda_k(t)|^2$. Applying this result in Eq.(\ref{psi1}) we obtain
\begin{eqnarray}\label{lado esquerdo}
\frac{d}{dt}\lvert \psi(t)\rangle_1 = \Big[\dot{\alpha}(t)\hat{a}^{\dagger} + \sum_{k}^{}\dot{\lambda}_k(t)\hat{b}_k^{\dagger}\Big]\lvert\psi(t)\rangle_1\ .
\end{eqnarray}
On the other hand
\begin{equation}\label{lado direito}
H_t\lvert \psi(t)\rangle_1 = \Big\{\omega_x + \Big[\big(\omega_c + \omega\big)\alpha(t) + \sum_{k}^{}\gamma_k\lambda_k(t)\Big]\hat{a}^{\dagger} + \sum_{k}^{}\Big(\omega_k\lambda_k(t) + \gamma_k\alpha(t)\Big)\hat{b}_k^{\dagger}\Big\}\lvert\psi(t)\rangle_1\ .
\end{equation}
Using Eq.(\ref{lado esquerdo}) and Eq.(\ref{lado direito}) on the Schr\"odinger equation, we obtain the following system of coupled equations for the amplitudes of the coherent states
\begin{eqnarray}\label{sistema 1}
i\dot{\alpha}(t) &=& \big(\omega_c + \omega\big)\alpha(t) + \sum_{k}^{}\gamma_k\lambda_k(t)\ ,\\\label{sistema 2}
i\dot{\lambda}_k(t) &=& \omega_k\lambda_k(t) + \gamma_k\alpha(t)\ .
\end{eqnarray}

By making the same calculation considering the second part of Eq.(\ref{estado evoluido t}), we obtain a similar set of coupled equations for $\beta(t)$ and $\chi(t)$ given by:
\begin{eqnarray}\label{sistema 3}
i\dot{\beta}(t) &=& \big(\omega_c - \omega\big)\beta(t) + \sum_{k}^{}\gamma_k\chi_k(t)\ ,\\\label{sistema 4}
i\dot{\chi}_k(t) &=& \omega_k\chi_k(t) + \gamma_k\beta(t)\ .
\end{eqnarray}

At this point, it is important to notice that the evolved state of the Eq.(\ref{estado evoluido t}), which is a result of the dispersive dynamics, is the initial input to generate Schr\"odinger cats inside the nanocavity.  Consider the coupling of the two-level system with an external probe: 
\begin{equation}\label{hamiltoniano rotacao}
\hat{H}_{\mathrm{probe}}=\frac{\Omega}{2}\left[e^{-i(\omega_p-\omega_x)\tau}\hat{\sigma}_+ + e^{i(\omega_p-\omega_x)\tau}\hat{\sigma}_-\right],
\end{equation}
where $\Omega$ represents the Rabi frequency describing the interaction of the laser with the qubit, and $\omega_p$ is the frequency of the pulsed laser. In order to eliminate the temporal dependence of the Hamiltonian, we will consider $\omega_p=\omega_x$, that is, a resonance condition between the external source and the qubit. If a pulse has a duration given by $\tau=\pi/2\Omega$ the two-level system is rotated, placing the field states in a quantum superposition.

\begin{eqnarray}\nonumber
\lvert\psi(t+\pi/2\Omega)\rangle&=&\left(\frac{\sin(\varphi)\lvert\alpha(t)\rangle\otimes\lvert\lambda(t)\rangle-i\cos(\varphi)\lvert\beta(t)\rangle\otimes\lvert\chi(t)\rangle}{\sqrt{2}}\right)\lvert1\rangle \\\label{superposicao mesoscopica}
& &+ \left(\frac{-i\sin(\varphi)\lvert\alpha(t)\rangle\otimes\lvert\lambda(t)\rangle + \cos(\varphi)\lvert\beta(t)\rangle\otimes\lvert\chi(t)\rangle}{\sqrt{2}}\right)\lvert0\rangle\ .
\end{eqnarray}
As can be seen, the net effect of the action of the pulse is to couple the excitonic states $\lvert 1\rangle$ and $\lvert 0\rangle$ in a mesoscopic superposition in the bosonic cavity, with terms in agreement with the form of Eq.(\ref{estado evoluido t}).

\subsection{Entanglement properties of system and reservoir}
\label{subsec:entanglement}
\hspace{10pt}In this section, we explore the characteristics of quantum information pertaining to two types of reservoirs coupled with the cavity. To examine the properties of quantum information in the interaction between the physical system and a reservoir, we employ the concept of quantum mutual information, which is defined as
\begin{equation}
\label{IMQ}
  I_{A:B}\equiv S(\rho_A)+S(\rho_B)-S(\rho_{AB}),  
\end{equation}
where $\rho_i$ is the density matrix of the system or subsystem and $S(\rho_i)$ is the von Neumann entropy.

About the reservoirs, the first type is a finite one, from which we extract a fragment to search for a quantifier of the information redundancy concerning the physical system in the environment. The second type is a structured reservoir that emerges in the context of solid-state devices, where interactions with the quantum system under study are not uniform across all frequencies. Instead, the environmental influence on the system varies significantly with frequency, which is modeled through a spectral density. This leads to nontrivial effects on the dynamics of the quantum system, particularly non-Markovian effects.

\subsubsection{Finite Reservoir}
\label{subsubsec:finiteR}
\hspace{10pt}If the qubit is in the state $\lvert 1\rangle$ with initial phase $\varphi=\pi/4$, The evolved state of the field and the reservoir will be in a superposition state, written as:
\begin{eqnarray}
\lvert\psi(t)\rangle_1 = \frac{\lvert \alpha(t)\rangle\otimes\lvert  \lambda(t)\rangle-i\lvert  \beta( t)\rangle\otimes\lvert  \chi(t)\rangle}{2}\ .
\end{eqnarray}
Therefore, the density operator for the combined field and reservoir system is:
\begin{eqnarray}\nonumber
\hat{\rho}_1(t) &=& \frac{1}{4}\Big\{\lvert   \alpha(t)\rangle \langle \alpha(t)\rvert \otimes\lvert   \lambda(t)\rangle \langle \lambda(t)\rvert  + \lvert\beta(t)\rangle \langle \beta(t)\rvert \otimes\lvert   \chi(t)\rangle\langle \chi(t)\rvert \\
& & + \big[i\lvert   \alpha(t)\rangle \langle \beta(t)\rvert \otimes\lvert   \lambda(t)\rangle \langle \chi(t)\rvert  + \mathrm{h.c.}\big] \Big\}\ .
\end{eqnarray}
 
Now, it is possible to calculate the reduced density operator belonging to each subsystem. For example, for the field inside the cavity is written as:
\begin{eqnarray}\nonumber
\hat{\rho}_c(t) &=& \mathrm{tr}_R[\hat{\rho}_1(t)]\\
&=& \frac{1}{4}\Big\{\lvert   \alpha(t)\rangle \langle \alpha(t)\rvert  + \lvert   \beta(t)\rangle \langle \beta(t)\rvert  + \big[i\langle \chi(t)|\lambda(t)\rangle \lvert   \alpha(t)\rangle \langle \beta(t)\rvert  + \mathrm{h.c.}\big]\Big\}\ .
\end{eqnarray}
As initially proposed by Blume-Kohout and Zurek \cite{blume2005}, we will focus on a specific region within the reservoir, referred to as the $F$ fragment. Considering that the system interacts with the degrees of freedom of $F$, for such a fragment of the reservoir $R$, we have:
\begin{eqnarray}\nonumber
\hat{\rho}_F(t) &=& \mathrm{tr}_{c,R-F}[\hat{\rho}_1(t)]\\ \nonumber
&=& \frac{1}{4}\left\{ {\lvert\lambda(t)\rangle}{\langle \lambda(t)\rvert_{F} } + {\lvert \chi(t)\rangle}\langle \chi(t)\rvert_{F} \right.\\
&& \left.+\left[i\langle \beta(t)\vert\alpha(t)\rangle_{R-F} {\langle \chi(t)\vert\lambda(t) \rangle}_{R-F} \lvert   \lambda(t)\rangle\langle \chi(t)\rvert_{F}  + \mathrm{h.c.}\right]\right\}\ .
\end{eqnarray}
Finally, the density operator for the combined cavity and fragment system is given by:
\begin{eqnarray}\nonumber
\hat{\rho}_{cF}(t) &=& \mathrm{tr}_{R-F}[\hat{\rho}_1(t)]\\\nonumber
&=& \frac{1}{4}\left\{\lvert\alpha(t)\rangle\langle \alpha(t)\rvert \otimes{\lvert\lambda(t)\rangle}{\langle \lambda(t)\rvert}_{F} + \lvert\beta(t)\rangle \langle \beta(t)\rvert \otimes{\lvert\chi(t)\rangle}{\langle \chi(t)\rvert_F }\right.\\ \nonumber
& & \left.+ \left[i{\langle \chi(t)|\lambda(t)\rangle}_{R-F} \lvert   \alpha(t)\rangle \langle \beta(t)\rvert \otimes\lvert   \lambda(t)\rangle\langle \chi(t)\rvert_{F}  + \mathrm{h.c.}\right]\right\}\ .
\end{eqnarray}
where  $\lvert \lambda(t)\rangle = \prod_{k}^{}\lvert \lambda_k(t)\rangle$, $\lvert \chi(t)\rangle = \prod_{k}^{}\lvert \chi_k(t)\rangle$, and the scalar product is given by $\langle \lambda(t)\vert\chi(t)\rangle = \prod_{k}\langle \lambda_k(t)|\chi_k(t)\rangle = \prod_{k} e^{-\frac{1}{2}(\lvert\lambda_k\rvert^2 + \lvert\chi_k\rvert^2) + \chi_k\lambda_k^{*}} = e^{\sum_{k}\{-\frac{1}{2}(\lvert\lambda_k\rvert^2 + \lvert\chi_k\rvert^2) + \chi_k\lambda_k^{*}\}}$.

\begin{figure}[ht]
	\centering
	\includegraphics[scale=0.25]{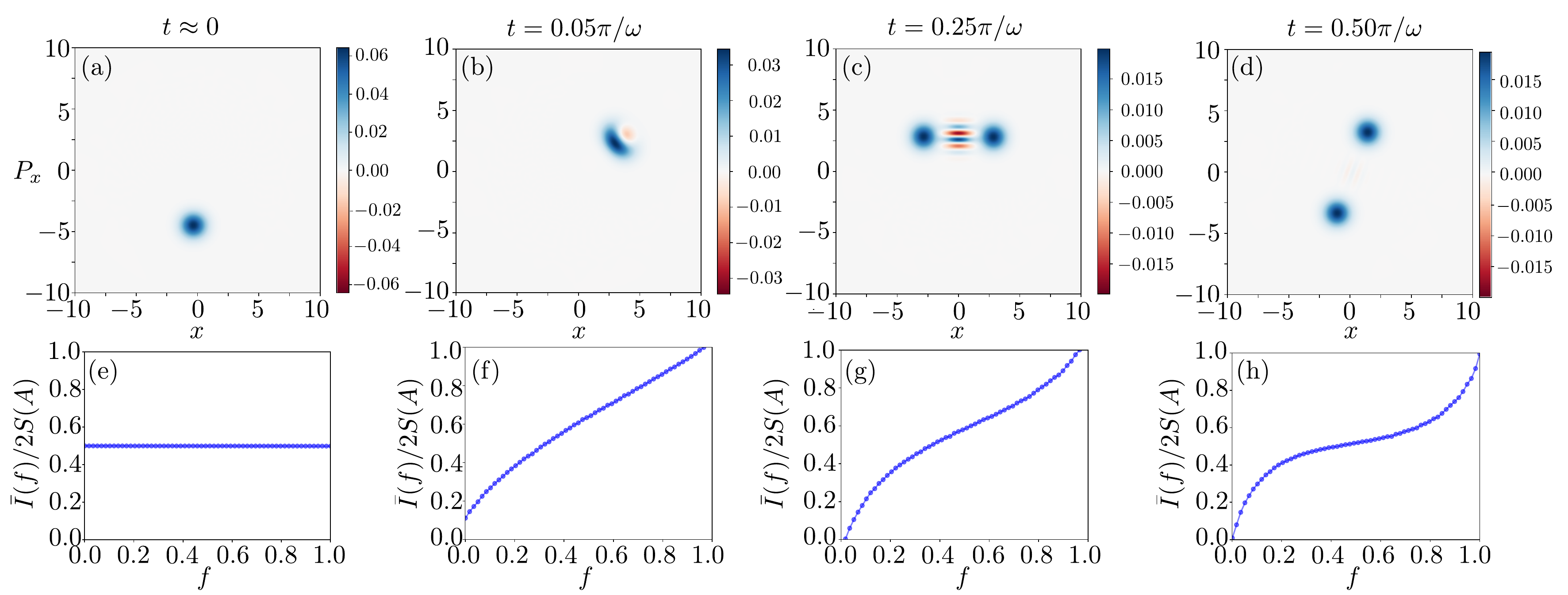}
	\caption{Comparison between the behavior of the Wigner function of the field inside the cavity, with the normalized average mutual information, $\bar{I}(f)/2S(A)$, for different instants of time $t$, with average number of photons $\lvert\alpha\rvert^2=10$. Here $f$ is the size fraction of the reservoir.}
	\label{gato-informacao}
\end{figure}

Quantum Mutual Information (QMI), $I_{A:F}$, as defined in Eq.(\ref{IMQ}), can be used to ascertain what is known about the system by the environment fragment. When calculating the von-Neumann entropies $S(A)$, $S(F)$ and $S(AF)$ of the system $A$, the fragment $F$ and the set $A+F$, respectively, $I_{A:F}$ can be obtained. Note that $I_{A:F} = 0$, if and only if, $\hat{\rho}_{AF} = \hat{\rho}_A\otimes\hat{\rho}_F$, that is, the subsystems are uncorrelated. When $\hat{\rho}_{AF}$ is a pure state, $I_{A:F} = 2 S(A)$, since $S(F) = S(A)$ and $S (AF) = 0$. To identify redundant correlations, we examine the partial information provided about all random fragments $F$ that contain a size fraction $f$ of $R$, such that $F = f N$, where $0\leq f \leq 1$ is the fraction of the environment that is taken. Therefore, the average mutual information over every fragment $F$ of size $f$ can be written as~\cite{blume2008}
\begin{eqnarray}\label{mutual information}
\bar{I}(f) = \left\langle I_{A:F}\right\rangle_F = \mathrm{avg}_{\mathrm{All\,}F\mathrm{\,of}\,\mathrm{size\,}f}(I_{A:F})\ .
\end{eqnarray}
In Fig.~\ref{gato-informacao}, the graphs of the Wigner function illustrate the formation and decay of a Schr\"odinger cat-like state in phase space, as shown in panels (a-d). The Wigner functions are associated with the coherent superposition of states of the qubit $\lvert 1\rangle$ and $\lvert 0\rangle$ from Eq.(\ref{superposicao mesoscopica}). We also present the normalized average mutual information (NAMI), $\bar{I}(f)//2S(A)$, which quantifies the information redundancy of the current system in the environment, panels (e-h). For our simulations, we consider $N=900$ as the number of oscillators in the reservoir. Within this limit, the dynamics of the finite reservoir  align with our numerical simulations, where the frequencies $\omega_k$ vary from $0$ to $10$, and the coupling constant is given by $\gamma_k=\omega/8$. The initial average number of photons is, $\lvert\alpha\rvert^2=\lvert\beta\rvert^2=10$ and and each simulation is averaged over 100 realizations. 

The Wigner function of the initial state of the nanocavity, as shown in Fig.~\ref{gato-informacao}(a), exhibits the characteristics of a Gaussian package with minimum uncertainty, indicative of a coherent state. At this moment, since all parts of the physical system are uncorrelated, the NAMI curve, Fig.~\ref{gato-informacao}(e), maintains a consistent value across any fraction size $f$. Upon evolution, the state in the cavity displays quantum signatures, evidenced by the emergence of squeezing in Fig.~\ref{gato-informacao}(b) at $t=0.05\pi/\omega$. Concurrently, if we look at Fig.~\ref{gato-informacao}(f), is observed an abrupt increment on the NAMI value. This behavior indicates that from the moment we switch on the interaction of the system with the reservoir, the mutual information increases, showing that the whole system (qubit, nanocavity and reservoir) moves from an uncorrelated state to a pure quantum state.

In Fig.\ref{gato-informacao}(c) and Fig.\ref{gato-informacao}(d), we observe the creation and disappearance of a coherent superposition of two mesoscopic states, representing the life and death of a Schr\"odinger cat-like state, respectively. Although the Wigner function in Fig.\ref{gato-informacao}(c) exhibits a negative part, the NAMI at this instant, as shown in Fig.\ref{gato-informacao}(g), indicates a correlation between the system's loss of coherence and the emergence of a plateau in mutual information. Over time, as more parts of the environment interact with the quantum system, it begins to lose coherence to the reservoir, leading to the formation of a plateau in mutual information. This plateau becomes increasingly pronounced when the cat state transitions into a statistical mixture, as depicted in Fig.\ref{gato-informacao}(d) with NAMI approaching $\bar{I}(f)/2S(A) = 0.5$ corresponding to an increase in the system's loss of coherence, as seen in Fig.\ref{gato-informacao}(h). 

This behavior can be interpreted as a transition point to classical behavior, where the quantum system begins to exhibit more deterministic characteristics and becomes less dependent on its quantum nature. Different sizes of fragments have access to the same amount of information about the system, so it is not necessary to monitor much of the environment to gather most of the information about the system. However, accessing such information becomes difficult since it is a large and complex system.

\subsubsection{Structured Reservoir}
\label{subsubsec:structuredR}
\hspace{10pt}We now return to Eqs.(\ref{sistema 1}) and making the substitution $\tilde{\alpha}(t) = \alpha(t)e^{i\Delta\omega t}$ and $\tilde{\lambda}_k(t) = \lambda_k(t)e^ {i\omega_k t}$, where $\Delta\omega = \omega_c + \omega$, we write
\begin{eqnarray}\label{sis 1}
i\dot{\tilde{\alpha}}(t) &=& \sum_{k}\gamma_k\tilde{\lambda}_k(t)e^{i(\Delta\omega - \omega_k) t}\\\label{sis 2}
i\dot{\tilde{\lambda}}_k(t) &=& \gamma_k\tilde{\alpha}(t)e^{-i(\Delta\omega - \omega_k)t}\ .
\end{eqnarray}
Integrating (\ref{sis 2}) and substituting in (\ref{sis 1}), we obtain
\begin{equation}
\dot{\tilde{\alpha}}(t) =-\sum_{k}\gamma_k^2\int_{0}^{t}\tilde{\alpha}(t')e^ {i(\Delta\omega - \omega_k)(t - t')}dt'
= -\int_{0}^{t}dt'f(t-t')\tilde{\alpha}(t')\ ,
\label{eq alpha}
\end{equation}
where we define Kernel memory
\begin{eqnarray}\label{kernel}
f(t-t') = \sum_{k}^{}\gamma_k^2 e^{i(\Delta\omega - \omega_k)(t - t')}\ .
\end{eqnarray}
The non-Markovian character of the problem is evident in the non-local time integral of the equation (\ref{eq alpha}). If the number of oscillators tends to infinity and a constant spectral density is assumed, the Kernel memory is approximately
\begin{eqnarray}
f(t-t') \approx 2\kappa\delta(t-t')\ ,
\end{eqnarray}
where $\kappa$ is a constant. In this case
\begin{eqnarray}
\dot{\tilde{\alpha}}(t) = -\int_{0}^{t}dt' 2\kappa\delta(t-t')\tilde{\alpha}(t') = -2 \kappa\tilde{\alpha}(t)\ ,
\end{eqnarray}
this equation has as a solution
\begin{eqnarray}\label{alpha markov}
\tilde{\alpha}(t) = \tilde{\alpha}(0)e^{-\kappa t}\ .
\end{eqnarray}
This is the well-known solution of the time evolution of the master equation in the Markov approximation, of the mode of a cavity that is initially in a given coherent state $\tilde{\alpha}(0)$. In terms of continuous limits of the ambient frequency, the correlation function, i.e., Kernel memory becomes
\begin{eqnarray}\label{kerne continua}
f(t-t') = \int_{0}^{\infty}J(\omega_k)e^{i(\Delta\omega - \omega_k)(t-t')}d\omega_k\ ,
\end{eqnarray}
Here, $J(\omega_k)$ represents the spectral density of reservoir modes. We assume a Lorentzian spectral density, which is typical of a structured cavity~\cite{Breuerbook}, whose shape is
\begin{eqnarray}
J(\omega_k) = \frac{1}{2\pi}\frac{\gamma\Lambda^2}{(\omega_c - \omega_k)^2 + \Lambda^2},
\end{eqnarray}
where parameter $\gamma$ is related to the microscopic system-reservoir coupling constant, while $\Lambda$ defines the spectral width of the cavity modes. It is worth mentioning that the parameters $\gamma$ and $\Lambda$ are related, respectively, to the correlation time of the reservoir $\tau_r$ and the relaxation time of the system $\tau_q$, with $\tau_r = \Lambda^{ -1}$ and $\tau_q \approx \gamma^{-1}$ \cite{Breuerbook}. The weak qubit-cavity coupling limit occurs for $\Lambda > \gamma$ ($\tau_r < \tau_q$). The inverse corresponds to the strong coupling limit $\Lambda < \gamma$ ($\tau_r > \tau_q$). The higher the cavity quality factor, the smaller the $\Lambda$ spectral width and, consequently, the photon decay rate. With this spectral density and applying the residue theorem~\cite{butkov1988}, the Kernel memory integral of Eq. (\ref{kerne continua}) can be solved, obtaining
\begin{eqnarray}
f(t-t') = \frac{\gamma \Lambda}{2}e^{-(\Lambda - i\omega)(t-t')}\ .
\end{eqnarray}
Substituting the result into the equation (\ref{eq alpha}) and solving using the Laplace transform, we obtain
\begin{eqnarray}\label{alpha t}
\tilde{\alpha}(t)=\tilde{\alpha}(0)\bigg[\cosh(\eta t/2) + \frac{M\sinh(\eta t/2)}{\eta} \bigg]e^{-\frac{1}{2}M t}\ ,
\end{eqnarray}
where $M = \Lambda - i\omega$ and $\eta = \sqrt{-2\Lambda\gamma + M^2}$. Thus, substituting (\ref{alpha t}) in (\ref{sis 2}) and integrating, we find the amplitude of the state belonging to the reservoir
\begin{eqnarray}\nonumber
\tilde{\lambda}_k (t) &=& \frac{i\gamma_k\tilde{\alpha}(0)}{-\eta^3 + 4\eta x^2}\Big\{2\eta [\cosh(x t) + \sinh(x t)](M-2x)\cosh(\eta t/2) \\
& &-4(Mx - \eta^2/2)[\cosh(x t) + \sinh(x t)]\sinh(\eta t/2) - 2\eta(M -2x)\Big\}\ ,
\end{eqnarray}
where $x = -(M/2 + i(\Delta\omega - \omega_k))$. Thus, making the inverse transformations $\alpha(t) = \tilde{\alpha}(t)e^{-i\Delta\omega t}$ and $\lambda_k(t) = \tilde{\lambda}_k (t)e^{-i\omega_k t}$, and performing analogous operations for $\beta(t)$ and $\chi(t)$, which are solutions of Eqs.(\ref{sistema 3}), we obtain the exact form of the evolved state $\left| \psi(t)\right\rangle$, Eq.(\ref{estado evoluido t}).

\begin{figure}[ht]
	\centering
	\includegraphics[scale=0.50]{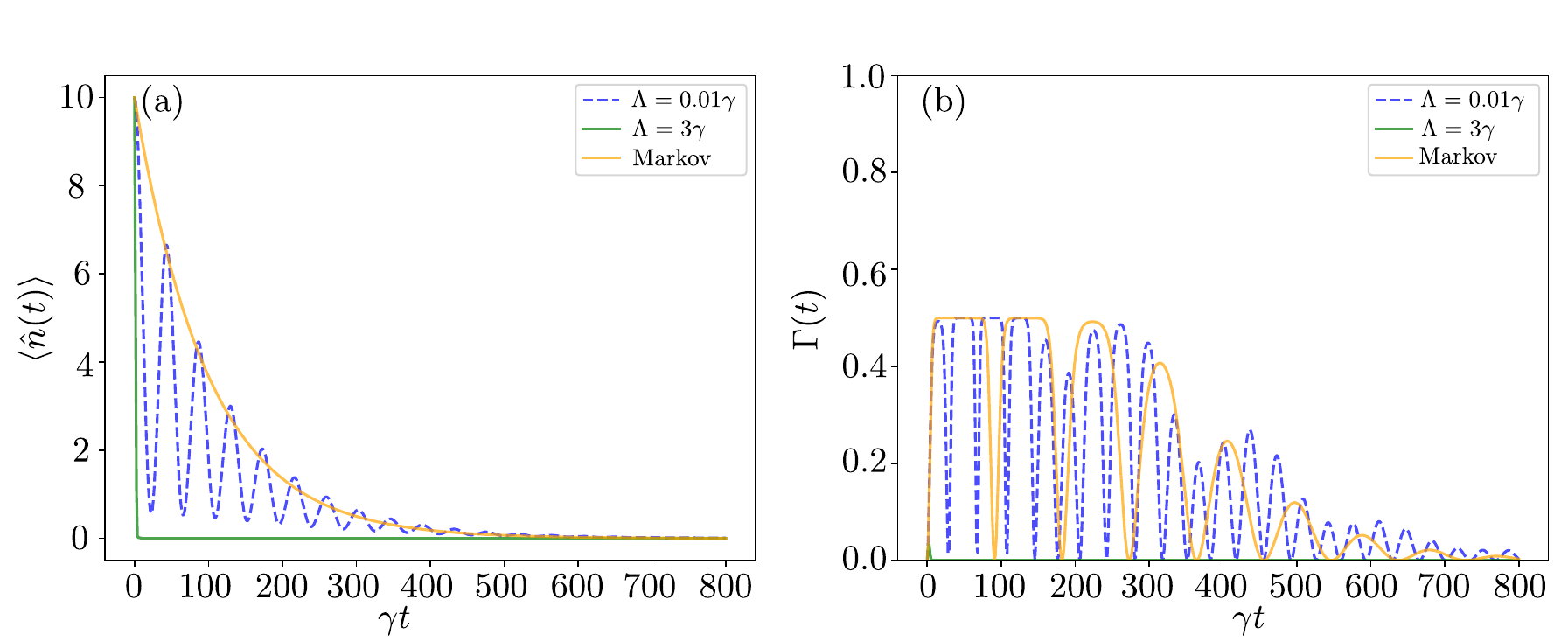}
	\caption{(a) Average number of photons $\left\langle \hat{n}(t)\right\rangle$ and (b) idempotency defect $\Gamma(t)$, as a function of time $\gamma t$. For average numbers of photons $|\alpha|^2 = |\beta|^2 = 10$. The limits are taken for analysis: $\Lambda=0.01\gamma$ in dashed blue line, equivalent to the strong coupling limit and $\Lambda=3\gamma$ in solid green line, modeling the weak coupling limit. For comparison purposes, the Markov limit is shown in the orange line.}
	\label{n-NM}
\end{figure}

In Fig.~\ref{n-NM} we present in (a) the behavior of the average number of photons $\left\langle \hat{n}(t)\right\rangle$ inside the cavity and in (b) the idempotency defect $\Gamma(t)=1-\mathrm{Tr}\rho^2_c$ of the same, both as a function of time, for an average number of photons $|\alpha|^2 = |\beta|^2 = 10$. Two limits are compared with the Markovian case (represented with the orange line in the figures): the strong coupling limit considering $\Lambda=0.01\gamma$, dashed blue line in both panels, and the weak coupling limit with $\Lambda =3\gamma$, solid green line. From the analysis of the figures, we found that in the strong coupling regime, both the average number of photons and the idempotency defect present oscillations. Oscillations in the number of photons are related to the exchange of excitations between the system and the reservoir. In the case of weak coupling, since the reservoir correlation time is negligible compared to the system relaxation time, the average number of photons quickly goes to zero. The oscillations on the idempotency defect shows that the system and reservoir interact strongly, exchanging information along the dynamics. 

\begin{figure}[ht]
	\centering
	\includegraphics[scale=0.50]{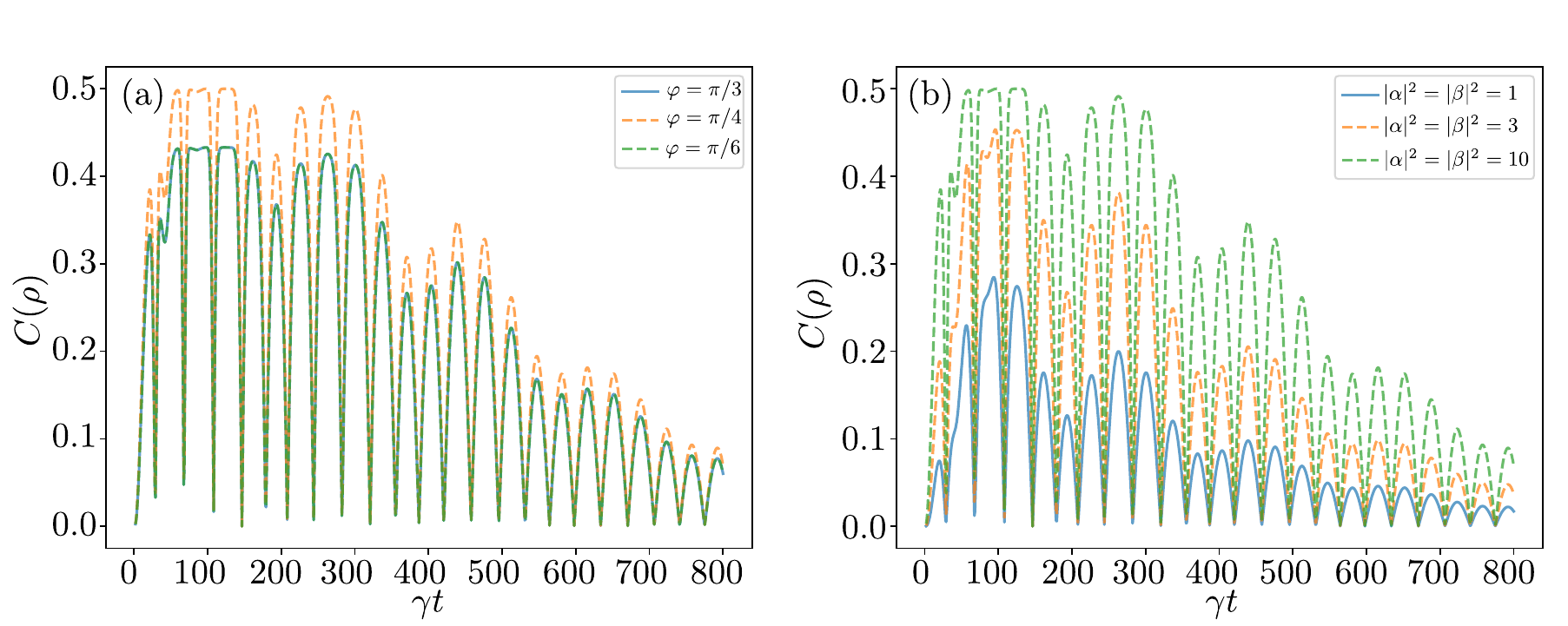}
	\caption{Concurrence dynamics $C(\rho)$ of the mesoscopic superposition state, Eq.(\ref{superposicao mesoscopica}), in the strong coupling limit $\Lambda=0.01\gamma$. In (a) for different weights between wave packets, broken down by the initial phase $\varphi$, with average initial photon number $|\alpha|^2 = |\beta|^2 = 10$. On the other hand, in graph (b) we observe the concurrence behavior for different average numbers of initial photons $|\alpha|^2 = |\beta|^2$, with initial phase $\varphi = \pi/4$.}
	\label{C}
\end{figure}

A special type of quantum correlation present in the system is entanglement, which has two sides: its notable side presents itself as a great resource for the development of quantum technologies, but it also has its villainous side, motivated through the decoherence process. Concerning our problem, at this point it becomes interesting to quantify the entanglement so we calculate the concurrence $C(\rho)$, to explore the degree of entanglement of the mesoscopic superposition state. With this goal, we use the transformation of the state of the products of coherent states into a base of two qubits, as described in Sec.\ref{sec:model}, Eqs.(\ref{eq:transformation2qubits}). On this new basis, the concurrence calculation can be carried out without major problems. We present our results in Fig.~\ref{C} for the strong coupling limit $\Lambda=0.01\gamma$, where panel (a) shows the behavior of concurrence as a function of time for different values of the phase $\varphi $ considering a fixed initial average number of photons $|\alpha|^2 = |\beta|^2 = 10$. In Fig.~\ref{C}(b) the behavior of concurrence for different average numbers of photons $|\alpha|^2 = |\beta|^2$ is studied, where it is evident that the entanglement-type correlations of the system is directly related to its size: where the larger it is, the more easily it becomes entangled with its environment. In contrast, the more entangled the system is with its environment, the faster it loses coherence.
As discussed, when the system is in the strong coupling limit the timescale of the system is comparable to that of the reservoir, so the reservoir is not fast enough to return to equilibrium and some memory effects accumulate. Such effects have a non-Markovian character and are taken into account by the non-local kernel time integral, Eq.(\ref{kernel}). 

\begin{figure}[ht]
	\centering
	\includegraphics[scale=0.50]{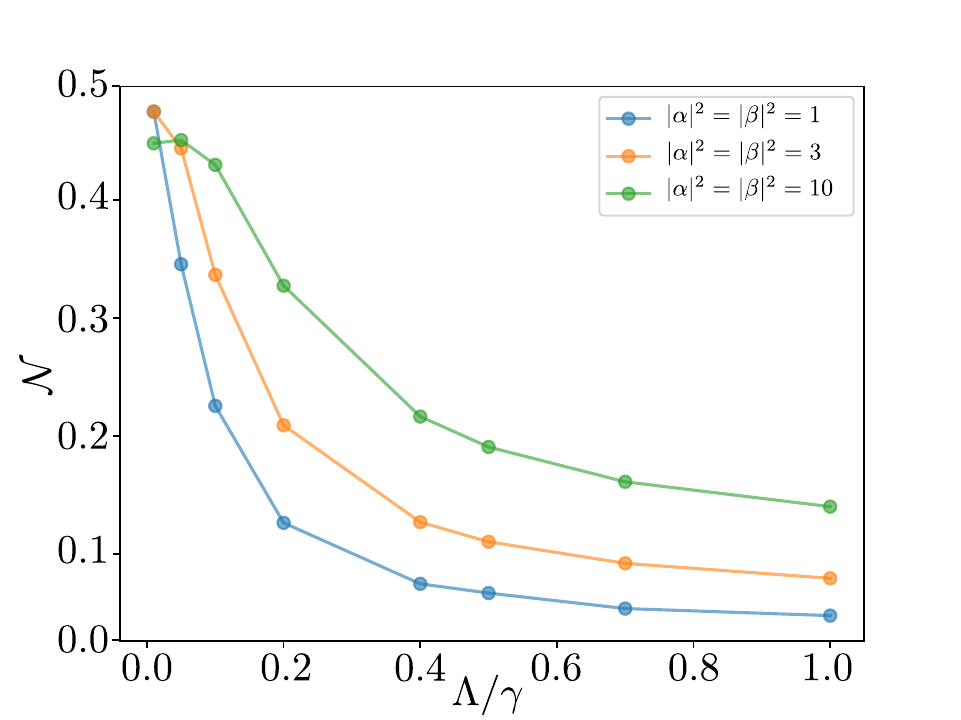}
	\caption{Measure of non-Markovianity of the system $\mathcal{N}$ as a function of $\Lambda/\gamma$, for different average numbers of initial photons $|\alpha|^2 = |\beta|^2$.}
	\label{NM}
\end{figure}

The commonly used approach for quantifying quantum non-Markovianity is based on the idea that memory effects in the dynamics of open systems are linked to the exchange of information between the open system and its environment. In this way, quantum non-Markovianity is associated with a notion of quantum memory, that is, information that was transferred to the environment, arising from system-environment correlations or changes in environmental states, and is subsequently recovered by the system. Based on this concept, we can introduce a well-established measure for the degree of memory effects~\cite{breuer,breuer2016}
\begin{eqnarray}\label{maximo}
\mathcal{N}=\max_{\rho^{1,2}_S}\int_{\sigma>0}^{}dt\sigma(t)\ ,
\end{eqnarray}
where
\begin{eqnarray}
\sigma(t)\equiv\frac{d}{dt}D[\rho^1(t),\rho^2(t)]
\end{eqnarray}
denotes the time derivative of the trace distance\footnote{It is an average of the distance between two quantum states $\rho_1(t)$ and $\rho_2(t)$, that is, similarity, defined as $D[\rho_1(t) ,\rho_2(t)] = (1/2)\mathrm{tr}|\rho_1(t) - \rho_2(t)|$, where $|X|=\sqrt{X^{\dagger}X} $.} of the pair of evolved states, which can be interpreted as characterizing a flow of information between the system and the environment. By construction, the integral is done over every interval in which $\sigma(t)>0$ and the maximization is done over all pairs of possible initial states. Therefore, $\mathcal{N}=0$, if and only if, the process is Markovian. It was shown that the pairs of optimal initial states $\rho^{1,2}_S$ lie on the boundary of the state space and, in particular, are always orthogonal \cite{wissmann2012}. In Fig.~\ref{NM} the measurement of $\mathcal{N}$ of the mesoscopic superposition state of two coherent wave packets (\ref{superposicao mesoscopica}) is presented as a function of the radius $\Lambda/\gamma $ for different average number of initial photons $|\alpha|^2 = |\beta|^2$. Note that, unlike the concurrence, as the average number of photons increases the degree of non-Markovianity of the system decreases, which is something of interest and requires a more rigorous investigation of this dynamic. Furthermore, large average numbers of photons have a lower degree of non-Markovianity initially, but they lose it more slowly compared to those with small average numbers of photons. However, as expected, they all always tend to zero in the weak coupling limit. These non-Markov effects will be investigated in more detail in future work.

\section{Summary}
\label{sec:summary}
\hspace{10pt}In this work, we propose the formation of a Schr\"odinger cat state within a system composed of an excitonic state coupled to a nanocavity, accounting for losses through two distinct types of reservoirs. The first, a finite reservoir, exhibits a plateau in the normalized average mutual information, signifying that reservoir fragments of varying sizes convey equivalent information about the physical system of interest. By transforming the reservoir into a structured one (an infinite number of oscillators with Lorentzian spectral density), we investigate non-Markovian effects on the cavity's photon count, idempotency defect, and calculate the concurrence between the reservoir and the system (via mapping to a two-qubit model), along with assessing memory effects. We observe that the dynamics in this latter scenario deviate significantly from the Markovian limit, exhibiting oscillations not only in the average photon count but also in the idempotency defect. These oscillations affect the information flow between the system and its environment, as evidenced by the non-Markovianity measures calculated.

\begin{acknowledgments}
This work was supported by CAPES, the Brazilian National Institute of Science and Technology of Quantum Information (INCT-IQ), grant 465469/2014-0/CNPq, and CNPq, grant 422350/2021-4.
\end{acknowledgments}

%

\end{document}